# Design of an all-fiber broadband mid-IR source through wavelength translation

A. Barh, S. Ghosh[#], R. K. Varshney and B. P. Pal*

Department of Physics, Indian Institute of Technology Delhi, New Delhi 110016, INDIA
[*]Tel: +91-11-26591327, Fax: +91-11-26581114, e-mail: bishnupal@gmail.com;
[#]INSPIRE-Faculty (DST-INDIA) Awardee-2012

**Abstract:** We report design of an efficient ~ 50 cm long all-fiber compact microstructured optical fiber-based 3-4.25 µm mid-IR light source with power conversion efficiency > 28% by exploiting FWM with $Er^{3+}$-doped ZBLAN fiber as the pump.

## 1. Introduction

Recent rapid advancement in fiber fabrication techniques and development of suitable relatively low-loss materials of good transparency have opened up a new platform for mid-IR photonics in the wavelength range 2 - 5 µm [1, 2]. Within this mid-IR window, wavelength range 3 - 4.2 µm is an attractive window for atmospheric transmission. This window can be explored to detect traces of environmental and toxic vapors as low as parts per billion for atmospheric, security and industrial applications. However, over the years it has been a challenge to develop simple but efficient scheme(s) to produce laser radiation to cover this relatively high-pass atmospheric transmission window. To meet this demand for a high-power, compact, efficient and reliable CW/ pulse laser sources in this mid-IR region, we propose an integrated all-fiber scheme exploiting four-wave mixing (FWM) for wavelength translation using the commercially available CW $Er^{3+}$-doped ZBLAN fiber as the pump ($\lambda_p$ ~ 2.8 µm). This superior FWM bandwidth (BW) was obtained through precise tailoring of the fiber's dispersion profile so as to realize positive quartic dispersion at the pump wavelength ($\lambda_p$). The fiber length was also optimized to ~ 0.5 m in order to achieve efficient phase matching between the propagating waves and the generated FWM signal.

## 2. Fiber design and wavelength translation

Among various nonlinear (NL) phenomena, FWM is the dominant mechanism for wavelength translation provided certain phase matching condition is satisfied. Under the degenerate FWM process, pump photons of frequency $\omega_p$ get converted to a signal photon ($\omega_s < \omega_p$) and an idler photon ($\omega_i > \omega_p$) according to the energy conservation relation ($2\omega_p = \omega_s + \omega_i$) where, subscripts s, i and p stands for signal, idler, and pump, respectively. In a highly NL single-mode fiber, the maximum frequency shift ($\Omega_s$) depends on both the magnitude and sign of its GVD parameters. On one hand, positive $\beta_4$ leads to broad-band and flat gain where as negative $\beta_4$ reduces the flatness and BW of FWM output. Thus higher order dispersion management is very crucial in such fiber designs to achieve a targeted output. This positive $\beta_4$ value in the vicinity of low negative $\beta_2$ could be achieved by suitably reducing the core cladding index difference ($\Delta n$). In our proposed fiber design, up to fourth order dispersion term is taken into account.

To achieve such application-specific fiber design, we focus on an arsenic sulphide ($As_2S_3$)-based microstructured optical fiber (MOF) geometry with a solid core and holey cladding, consisting of 4 rings of hexagonally arranged holes embedded in $As_2S_3$ matrix (Fig. 1(a)). In order to reduce $\Delta n$, we choose thermally compatible borosilicate glass rods to fill the holes. To limit confinement loss ($\alpha_c$ < 1dB/m) and to tune dispersion curve accordingly, we have chosen the sizes of rods in the $2^{nd}$ cladding ring (radius $r_2$) different from the surrounding rings. After optimization we have fixed fiber parameters as $d/\Lambda = 0.5$, $\Lambda = 2.5$ µm and $r_2 = 0.635$ µm. To attain sufficient signal amplification factor ($AF_s$), which is a measure of the achievable gain over the targeted wavelength regime of 3 ~ 4.2 µm, we have assumed 10 W of input pump power ($P_0$). In order to suppress other NL effects, short fiber length is considered and optimized to ~ 50 cm. For this fiber structure dispersion ($D$) and $\beta_2$ variation is shown in Fig. 1(b). It can be seen that the zero dispersion wavelength in the designed fiber is 2.788 µm.

In our design calculation, initially we have studied the FWM performance under lossless, undepleted pump condition, where $P_0$ is only transferred to $\lambda_s$ and $\lambda_i$. We have also shown that launching of a weak idler ($P_{I,in}$) along with the pump improves the FWM efficiency through stimulated FWM. Variation of $AF$ of output spectrum has been studied with different $\lambda_p$ (as shown in Fig. 1(c)), where the required output BW (3 – 4.25 µm) with sufficient amplification (~ 38 dB) is achieved for a suitable choice of pump operating near $\lambda_p = 2.792$ µm.

In the next step, assuming quasi-CW conditions, we have studied the complex amplitudes $A_j(z)$ (j = p, i, s) and corresponding power ($P_{out}$) variations along the fiber length ($L$) to study the effect of pump depletion and spectral dependence material loss ($\alpha_j$) by numerically solving the three coupled amplitude equations (Eq. 1 – 3) [4].

$$\frac{dA_p}{dz} = -\frac{\alpha_p A_p}{2} + \frac{in_2\omega_p}{c}\left[\left(f_{pp}|A_p|^2 + 2\sum_{k=i,s} f_{pk}|A_k|^2\right)A_p + 2f_{ppis}A_p^* A_i A_s e^{j\Delta k_L z}\right] \quad (1)$$

$$\frac{dA_i}{dz} = -\frac{\alpha_i A_i}{2} + \frac{in_2\omega_i}{c}\left[\left(f_{ii}|A_i|^2 + 2\sum_{k=p,s} f_{ik}|A_k|^2\right)A_i + f_{ispp}A_s^* A_p^2 e^{-j\Delta k_L z}\right] \quad (2)$$

$$\frac{dA_s}{dz} = -\frac{\alpha_s A_s}{2} + \frac{in_2\omega_s}{c}\left[\left(f_{ss}|A_s|^2 + 2\sum_{k=p,i} f_{sk}|A_k|^2\right)A_s + f_{sipp}A_i^* A_p^2 e^{-j\Delta k_L z}\right] \quad (3)$$

where, $n_2$ is the NL index coefficient and $\Delta k_L$ is the linear phase mismatch term, $f_{jk}$ and $f_{ijkl}$ are overlap integrals. We have fixed $P_0$ at 10 W and $\lambda_p$ at 2.792 µm and optimized $P_{I,in}$ as 20 mW. As a sample, for one set of $\lambda_s$ (4.13 µm) and $\lambda_i$ (2.11 µm) $AF$ and $P_{out}$ variations along $L$ were studied (shown in Fig. 1(d) & (e), respectively). Optimum $L$ becomes ~ 50 cm. Even after inclusion of pump depletion and loss, average $P_{s,out} > 2.83$ W is achievable with a conversion efficiency > 28%. The entire signal spectrum is shown in Fig. 1(f). It can be seen that the spectral BW is very similar to the undepleted case, the maximum $AF_s$ however decreases to ≈ 22 dB due to inclusion of pump depletion and loss. Such a fiber, if experimentally realized should be attractive as a mid-IR light source for a variety of applications.

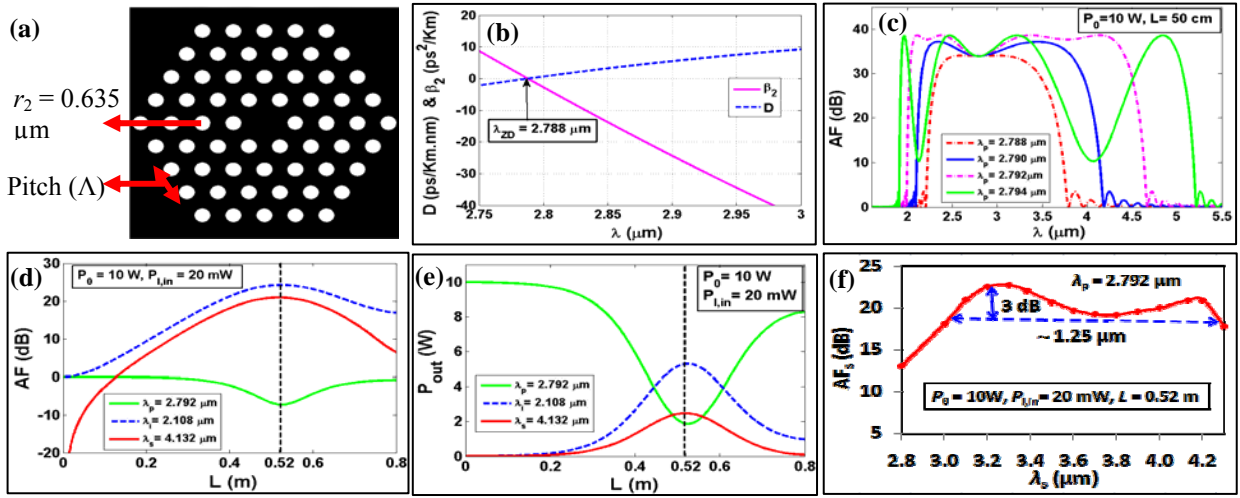

Fig. 1: (a) Cross section of the designed MOF. Cladding consists of 4 rings of borosilicate rods (white circles) embedded in the $As_2S_3$ matrix (black background); (b) Dispersion (blue dotted) and $\beta_2$ (pink solid) variation, $\lambda_{ZD} = 2.788$ µm; (c) Variation of $AF$ for different $\lambda_p$ neglecting pump depletion and loss. With pumping at 2.792 µm (pink dotted), output signal spectrum is almost uniform with 3-dB BW ranging from 3 - 4.25 µm; (d) Variation of $AF$ for $\lambda_P$, $\lambda_s$ and $\lambda_i$ along length including pump depletion and loss; (e) Output power ($P_{out}$) variation; (f) Optimum signal 3-dB BW (3 – 4.25 µm) with pump depletion and loss. Maximum $AF_s$ is ≈ 22 dB.

## 3. Conclusions and Acknowledgement

We report design of an all-fiber compact and efficient 3 - 4.25 µm light source based on a 50 cm long specialty MOF, in which achievable amplification is shown to be (> 20 dB) with a 2.79 µm pump of 10 W power. The achievable average power conversion efficiency is shown to be > 28%. Thus our proposed fiber-based broad-band source should be useful to explore the new avenues of research in mid-IR photonics for military, spectroscopy as well as astronomical applications.

Govind Agrawal of University of Rochester is thanked for his advice and useful comments on our earlier related work in this area. This work relates to Department of the Navy Grant N62909-10-1-7141 issued by Office of Naval Research Global. The United States Government has royalty-free license throughout the world in all copyrightable material contained herein.